\documentclass[3p,times,twocolumn]{elsarticle}

\usepackage{amssymb}
\usepackage[figuresright]{rotating}

\begin{document}

\begin{frontmatter}



%% e.g. \dochead{17th International Conference on Dynamical Processes in Excited States of Solids}

\title{Precise measurement of the $^{222}$Rn half-life: a probe to monitor the stability of radioactivity}


\author[mi]{E. Bellotti}
\author[pd]{C. Broggini\corref{cor1}}
%\ead{broggini@pd.infn.it}
\author[lngs]{G. Di Carlo}
\author[lngs]{M. Laubenstein}
\author[pd]{R. Menegazzo}

\address[mi]{Universit\`{a} degli Studi di Milano Bicocca and Istituto Nazionale di Fisica Nucleare, Sezione di Milano, Milano, Italy}
\address[pd]{Istituto Nazionale di Fisica Nucleare, Sezione di Padova, Padova, Italy}
\address[lngs]{Istituto Nazionale di Fisica Nucleare, Laboratori Nazionali del Gran Sasso, Assergi (AQ), Italy}

\cortext[cor1]{Corresponding author}

\begin{abstract}
We give the results of a study on the $^{222}$Rn  decay we performed in the Gran Sasso
Laboratory (LNGS) by detecting the gamma rays from the radon progeny.
The motivation was to monitor the stability of radioactivity measuring several times per year the half-life of a short lifetime (days) source
instead of measuring over a long period the activity of a long lifetime (tens or hundreds of years) source.
In particular, we give a possible reason of the large periodical fluctuations in the count rate of the gamma rays due to radon
inside a closed canister which has been described in literature and which has been attributed to a possible influence of a component in the solar irradiation affecting the nuclear decay rates. We then provide the result of four half-life measurements we performed underground at LNGS in the period from May 2014 to January 2015 with radon diffused into olive oil. Briefly, we did not measure any change of the $^{222}$Rn half-life with a 8$\cdot$10$^{-5}$  precision. Finally, we provide the most precise value for the $^{222}$Rn half-life: 3.82146(16)$_{stat}$(4)$_{ syst}$  days.
\end{abstract}

\end{frontmatter}

% \linenumbers

\section{Introduction}
A possible time dependence of the radioactive nuclei decay constant has been searched for since the beginning of the science of radioactivity.
For instance, in the Ph.D. of M. Curie \cite{Cur03} one can already find the description of the search for a difference in the radioactivity of uranium between midday and midnight.
Recently, in particular since the year 2009 \cite{Jen09}, various experiments have reported evidence of a time modulation of the  decay constant of several  radioactive nuclei
with period, in most cases, of one year (but also of about one month or one day) and amplitude at the per mil level. This annual modulation, with the maximum in February and the minimum in August, has been correlated to the change of the Sun-Earth distance between aphelion and perihelion; however different measurements exclude any modulation as large as the reported ones (for recent reviews on results with and without modulation \cite{Stu14,Bel14}).

In \cite{Jen09,Fis11} the existence of new and unknown particle interaction has been advocated to explain the yearly variation in the activity of radioactive sources, with the particle being emitted from the Sun.
In \cite{Bel14} the possibility that the coupling to a long range scalar field, sourced by the Sun, might be the origin of the modulation has been quantitatively discussed.
The laboratory constraints on the variation of $\alpha_{em}$ and of the electron to proton mass ratio on an annual timescale turned out to induce upper bounds to the relative variation of the decay constant nine orders of magnitude lower than the claimed per mil effect.
Solar neutrinos have also been proposed as responsible for the modulation, with a cross section several orders of magnitude higher than expected. In particular, the possibility for anti-neutrinos affecting the $\beta^{+}$ decay of $^{22}$Na has been very recently studied in a reactor experiment \cite{Mei14} .

Since we believe that dedicated experiments are still needed to clarify the somehow contradictory situation,
in this letter we describe the results of a different
approach we pursue to monitor the time dependence of
radioactivity.

\begin{table}[ht]
\caption{The $^{222}$Rn decay chain \cite{Fir99}. Gamma rays with relative probability smaller than 3$\%$ are omitted.}
\footnotesize
\begin{center}
\begin{tabular}{|c|c|c|c|c|}
\hline
Isotope & Decay & half-life & Gamma & Relative \\
& type & & energy [keV] & probability \\
\hline
$^{222}$Rn & $\alpha $ & 3.8 d  &     &    \\
$^{218}$Po & $\alpha $ & 3.1 m  &     &    \\
$^{214}$Pb & $\beta  $ & 26.8 m & 242 & 7$\%$  \\
 & & & 295 & 18$\%$  \\
 & & & 352 & 36$\%$  \\
$^{214}$Bi & $\beta  $ & 19.9 m & 609 & 45$\%$  \\
 & & & 768 & 5$\%$  \\
 & & & 934 & 3$\%$  \\
 & & & 1120 & 15$\%$  \\
 & & & 1238 & 6$\%$  \\
 & & & 1378 & 4$\%$  \\
 & & & 1764 & 15$\%$  \\
 & & & 2204 & 5$\%$  \\
$^{214}$Po & $\alpha $ & 164 $\mu$s  &     &    \\
$^{210}$Pb & $\beta  $ & 22.3 y & 46.5 & 4$\%$  \\
$^{210}$Po & $\alpha $ & 138 d &     &    \\
$^{206}$Pb & & & & \\
\hline
\end{tabular}
\end{center}
\end{table}

Generally, a long lifetime source (tens or hundreds of years) is selected and its activity
is measured for a period of at least one year. This way both the count rate and the dead time are almost constant.
We also followed this approach in the study of $^{137}$Cs \cite{Bel12}, $^{40}$K and $^{232}$Th \cite{Bel13,Bel14}. However,
during such a long period of time several things may change, such as the laboratory pressure, temperature and humidity, the radioactive background, the electronic noise, and the performances of the electronic chain. It is not easy to keep all these under control if looking for a variation at the per mil level, or even below.
To mitigate the effect of all these possible instabilities  we decided to measure time after time the half-life of a short lifetime (days) source and for this purpose we selected $^{222}$Rn, a source relatively easy to produce. A similar approach has already been followed in \cite{Har12} and
\cite{Ale14} with the measurement of the
$^{198}$Au and $^{214}$Po half-life, respectively, with a precision of a few parts over 10$^{4}$.

In the next section we describe the experiment we performed with radon in a confined air volume, before giving in section 3 the results of the measurements we performed
underground at LNGS
with radon diffused into olive oil.

\section{Radon in air}
$^{222}$Rn is a noble radioactive gas coming from the decay of $^{226}$Ra (half-life: 1600(7) y \cite{Fir99} ) in the $^{238}$U chain. The decay chain below $^{222}$Rn
(half-life: 3.8232(8) d \cite{Mon08}) is given in Table 1, together with the energy and the relative probability of the emitted gamma rays. In our experiment we measure the
$^{222}$Rn half-life with gamma spectroscopy: this way we are sensitive to variations both in alpha and beta decay.

\begin{figure}[t]
\vskip -5mm
\centerline{
\includegraphics[width=0.52\textwidth,angle=0]{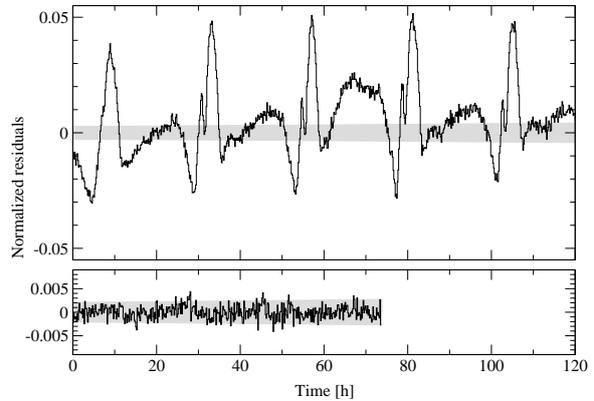}
}
\vskip -7mm
\caption{Normalized deviations from the exponential trend of the measured count rate as function of time for radon in air: without (upper panel) and with (lower panel) the  polystyrene particles inside the glass sphere containing air charged with radon. The shaded areas limit the region where fluctuations are within $\pm$2$\sigma$.}
\label{ampute}
\end{figure}

The set-up we used in the external Gran Sasso Laboratory during the preparatory phase of the experiment (to study the source production) is rather simple: a glass sphere (130 mm diameter) is connected through a pipe to a stainless steel cylinder containing a 0.3 kg rock rich in uranium. The radon from the radium decays in the rock fills by diffusion the glass sphere which, after 5-6 days, is isolated from the radon source by closing a Nupro (R) valve.

Gamma rays from radon progeny are detected by a 3"x3" NaI crystal placed at a few millimeter distance from the sphere surface
(the geometrical centers of the sphere and of the crystal are aligned).
The NaI is powered and read out through an Ortec (R) digiBASE (TM) which also pre-amplify and digitize the signal. Both the detector and the glass sphere are enclosed inside a 5 cm thick lead shield.
The detector count rate, i.e. the integral of the spectrum above 6 keV threshold, is of about 1000 counts per second (cps) at the beginning of the measurement, to be compared to a background, before filling with radon, of about 5 cps.

The normalized residuals of the count rate, i.e. the difference between the measured rate and the expected one, divided by the expected rate as function of time is shown in Figure 1 (upper panel). Instead of having a statistical distribution around zero, there is clearly a 24 hour period.
We have the same behavior if, instead of considering the whole energy spectrum, we take into account only the peaks
due to $^{214}$Pb and $^{214}$Bi decay. A similar effect has already been observed in the gamma radiation from radon progeny within air in confined conditions \cite{Stu12,Ste13,Ste14} and it
has been attributed to a possible influence of a component in the solar irradiation affecting the nuclear decay rates. In our opinion, all this could
\begin{figure}[t]
\vskip -5mm
\centerline{
\includegraphics[width=0.55\textwidth,angle=0]{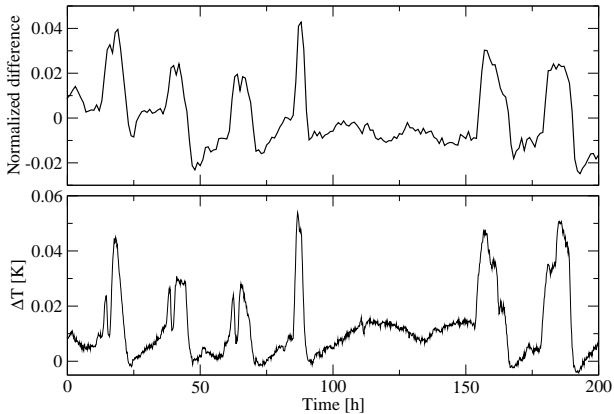}
}
\vskip -7mm
\caption{Normalized difference of detector count rates (D1-D2) as function of time with the two detectors placed on opposite sides of the glass sphere filled with radon (top panel). Typical 2$\sigma$ fluctuations amount to about  0.001. The time dependence of the temperature difference (T2-T1) measured on the sphere surface  is shown in the lower panel. The hours from 100 to 150 correspond to a week-end, when laboratory
temperature is very stable.}
\label{ampute}
\end{figure}
simply be attributed to the displacement of the radioactive nuclei inside the gas volume, with a variation in the detection efficiency.

We tried to verify our hypothesis by completely filling the glass sphere with spherical polystyrene particles (diameter: 0.7-0.9 mm). This way the atoms  cannot move freely within the entire volume but
they are essentially confined to the small volume within the interstitial space among polystyrene particles. The background, measured during 20 days before filling with radon, had still a stable rate of about 5 cps. The glass sphere was then charged with radon for two weeks to give an initial count rate of about 1500 cps. Spectra taken with this configuration do not exhibit any modulation, as shown in Figure 1 (lower panel).
As a consequence, we conclude that the displacement of radioactive atoms is the reason of the effect.

Searching for the trigger of this displacement, we modified the experimental set-up by adding a second 3"x3" NaI detector (the two detectors were symmetrically placed on opposite side of the glass sphere). In addition, we
measured the temperature at different places on the sphere by using precision integrated-circuit temperature sensors with output voltage linearly proportional to the Centigrade temperature. In particular, we measured the temperature in the two points on the surface just facing the two detectors.

We noticed that
the temperature change is not the same everywhere on the surface of the glass sphere. In addition, we had a clear correlation (Figure 2) between
the count rate difference of the two detectors D1 and D2 and the temperature difference between the two points  of the sphere in front of the detectors where the temperature T1 and T2 were measured (T1 and T2 were taken in front of D1 and D2, respectively). The count rate of the detector close to the higher temperature point was lower than the one of the detector close to the lower temperature point (the maximum temperature difference amounted to 0.05 K, while the maximum temperature excursion was from 13.5 to 17 $^{\circ}$C). Finally, we were able to reproduce the count rate modulation by switching on and off a heater placed at 1 m distance from the lead shield.

Our conclusion is that large changes in the count rate in addition to the exponential decay are due to the displacement of
radon and its progeny inside the glass sphere volume triggered by a small temperature gradient among different points  on the sphere surface (as shown in Figure 2, a difference of
0.01 K is enough to produce a sizable effect on the count rate).
The displacement of the radioactive atoms gives then rise to a different detection efficiency.

\section{Radon in liquid}
For a precise study of the $^{222}$Rn half-life it is necessary to suppress the temperature effect described in the previous section. Therefore, we selected an option able to 'immobilize' radon and its progeny but allowing at the same time for a rather high radon concentration and, as a consequence, a high count rate. As a matter of fact, a statistical analysis shows that a precision on the life time better than one part over 10$^{4}$ requires an initial count rate of at least 400 cps and a measuring time of about 2 months. To achieve this we diffused radon in olive oil which has much higher viscosity than air and which permits in equilibrium conditions a
radon concentration 29 times higher than in air \cite{Sol82}.

\begin{figure}[h]
\vskip -3mm
\centerline{
\includegraphics[width=0.52\textwidth,angle=0]{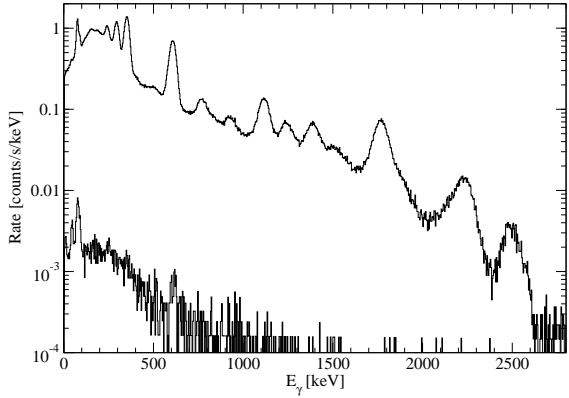}
}
\vskip -7mm
\caption{Spectra collected underground before charging the oil with radon (lower graph) and during a radon run (upper graph).
}
\label{ampute}
\end{figure}

For the half-life measurement we minimized the background and its fluctuations by running the experiment underground at LNGS, where the muon and neutron flux are suppressed by six and three orders of magnitude, respectively, and by having a copper cube (10 cm side, 2 mm thick walls) to contain the olive oil charged with radon. In particular, the copper cube filled with olive oil was charged with radon in the external laboratory by slowly pumping and diffusing
through it the air coming from the cylinder containing the uranium rich rock. After about one week the cube was
isolated, detached from the cylinder, hermetically closed and moved underground. The detector was a 3"x3" NaI powered and read out by a digiBASE (TM).  The shielding to absorb the external gamma rays was made of at least 15 cm of lead and the laboratory temperature was kept between 12 and 13 $^{\circ}$C.

\begin{figure}[h]
\vskip -3mm
\centerline{
\includegraphics[width=0.52\textwidth,angle=0]{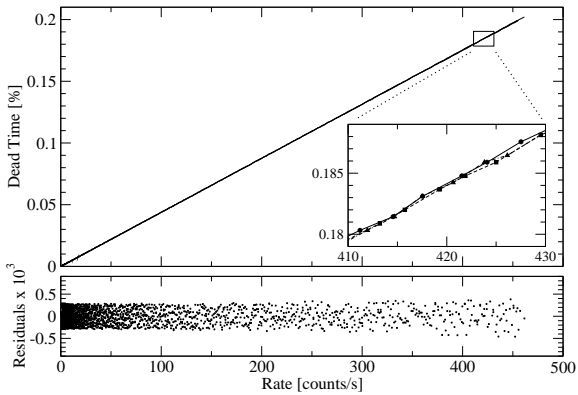}
}
\vskip -7mm
\caption{Dead time as function of count rate for the half-life measurements. A detailed view at the beginning of the runs is given in the insert. Absolute deviations from a linear dependence are shown in the lower panel.}
\label{ampute}
\end{figure}

Figure 3 shows the spectra collected underground before charging the oil with radon and during a radon run.
All the peaks due to the gamma rays given in Table 1 or to their sum are clearly visible.
In addition, there is a small bump at 511 keV  due to e$^{+}$e $^{-}$  annihilation from gamma rays interacting in the lead shield. In the analysis we consider
the entire energy spectrum above 6 keV and not only the full energy peaks. This because
we want to avoid any inaccuracy coming from the fitting procedure and we also want
to increase the total rate in order to improve the statistics.

Spectra are collected during consecutive time windows of 3600 s each, with
the timing provided by the internal quartz
oscillator of the digiBASE (TM). Its precision and stability
(better than 5 ppm/year) are enough for our purposes. The time stamp of each 3600 s measurement is then converted into the UTC (Coordinated Universal Time)
time scale using a periodic check of the internal time of the data acquisition PC (free running) with respect to the UTC itself. This conversion gives rise to a relative systematic uncertainty on the life time measurement of less than 2$\cdot$10$^{-6}$.

In the period from May 2014 to January 2015 we took four different measurements of the $^{222}$Rn activity to extract the half-life and to measure its stability.
Before charging with $^{222}$Rn we took a 20 day long background
run with the copper box filled with olive oil and closed inside the lead shield underground. We took this measurement to preliminary check the background stability over time.

\begin{table*}[ht]
\caption{The parameters from the fit A+B$\cdot$e$^{-t/C}$ to the count rate as function of time.}
\footnotesize
\begin{center}
\begin{tabular}{|c|c|c|c|c|c|}
\hline
Run & Running time [h] & A [cps] & B [cps] & C [d] & reduced  $\chi^2$ \\
\hline
1 & 1301 & 1.0209(9)  & 452.98(5) & 5.51335(46) & 1.13 \\
2 & 1462 & 1.0239(7)  & 457.98(5) & 5.51302(44) & 1.06 \\
3 & 1185 & 1.0247(10) & 463.16(5) & 5.51352(46) & 1.09 \\
4 & 1357 & 1.0241(9) & 354.82(4) & 5.51289(50) & 1.03 \\
\hline
\end{tabular}
\end{center}
\end{table*}

Dead time, provided by the digiBASE (TM) with 20 ms resolution,  amounts to 0.2$\%$ at the beginning of the radon measurement, linearly decreasing with the count rate, as shown in Figure 4. From the lower panel in Fig. 4 we see
that, for a constant count rate, there is a maximum relative fluctuation of the dead time of 4$\cdot$10$^{-3}$. As a consequence, the maximum
relative systematic uncertainty on the life time measurement due to dead time fluctuations amounts to at most 8$\cdot$10$^{-6}$, one order of magnitude smaller than the statistical error.

\begin{figure}[h]
\vskip -3mm
\centerline{
\includegraphics[width=0.52\textwidth,angle=0]{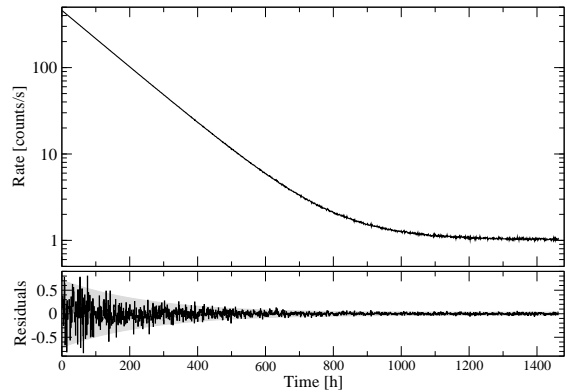}
}
\vskip -7mm
\caption{Count rate as function of time and normalized residuals for the second half-life measurement. The shaded area corresponds to the $\pm$2$\sigma$ uncertainty region.}
\label{ampute}
\end{figure}

The count rate as function of time and the normalized residuals for the second measurement are shown in Figure 5 (for the other three measurements the figures are essentially the same). For each measurement we fit the count rate, after dead time correction, with the function A+B$\cdot$e$^{-t/C}$. In Table 2 the values of the parameters and of the reduced chi squared  are given, with
$\chi^2$ computed by taking into account only statistical uncertainties from the Poisson distribution.

The fit parameter A corresponds to the background which, as shown in Table 2, is slightly increasing during the measurements. Figure 6 shows the lowest energy part of the background spectrum taken in one hour before starting the radon measurement and during the last hour of the third radon measurement, when radon has completely decayed. It is interesting to note that the 46.5 keV peak due to $^{210}$Pb, which is the background component expected to increase with time during the radon measurement because of its long half life, has increased by less than 3$\cdot$10$^{-3}$
cps (the 46.5 keV peak in the background spectrum before filling with radon is due to $^{210}$Pb decay in the lead shield). If we impose into the fit a slope on the background of 10$^{-3}$ cps/60 days we then have a relative variation of the half life of 7$\cdot$10$^{-6}$, one order of magnitude smaller than the statistical error.

\begin{figure}[h]
\vskip -3mm
\centerline{
\includegraphics[width=0.52\textwidth,angle=0]{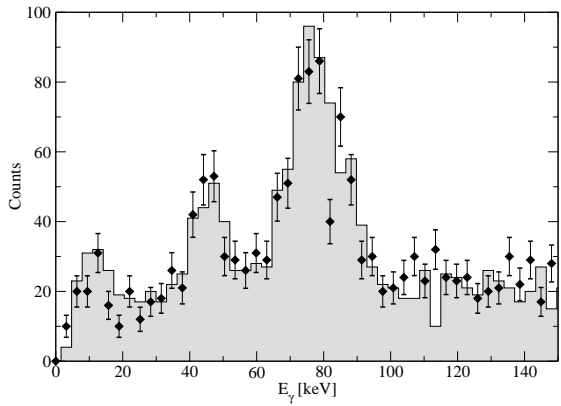}
}
\vskip -7mm
\caption{The lowest energy part of the background spectrum taken in one hour before starting the radon measurement (shaded area) and during the last hour of the third radon measurement (points). Statistical uncertainties are shown only for the radon measurement. In addition to the 46.5 keV peak  from $^{210}$Pb decay, the $L$ (around 12 keV) and $K$ (around 75 keV) X-rays of lead are clearly visible.
}
\label{ampute}
\end{figure}

From the four measurements we obtain the following values for the half-life:
3.82157(32) d, 3.82134(30)
d, 3.82169(32)  d and 3.82124(35).
They are perfectly compatible and, as a consequence, the $^{222}$Rn half-life has been constant, within 8 parts over  10$^{5}$ (1 sigma error for each measurement) during 9 months from May 2014 to January 2015.
We are now planning to continue these half-life measurements for at least
another year.

Finally, if we average the results of the four independent and mutually consistent measurements we obtain the most precise value for $^{222}$Rn half-life: 3.82146(16)$_{stat}$(4)$_{ syst}$  d, where the value of the systematic uncertainty is the sum in quadrature of all the previously described systematic uncertainties which are given in Table 3. Our value has to be compared to the most precise recent measurement of 3.8224(18) d, obtained with a $^{222}$Rn aqueous solution in liquid scintillator \cite{Col95}, and to the world average of 3.8232(8) d \cite{Mon08}. We observe that the world average is a weighted average of seven measurements \cite{Chi08}, several of which are very old, with limited description of the systematics but with very small error, much smaller than in \cite{Col95}.
\begin{table}[ht]
\caption{The relative systematic uncertainties on the $^{222}$Rn half-life (in 10$^{-6}$ units).}
\footnotesize
\begin{center}
\begin{tabular}{|c|c|c|c|}
\hline
Time conversion & Dead time fluctuations & $^{210}$Pb & Total  \\
\hline
2 & 8 & 7  & 10.8 \\
\hline
\end{tabular}
\end{center}
\end{table}

\section{Conclusions}
We have studied the decay of $^{222}$Rn at the Gran Sasso
Laboratory by detecting the gamma rays from its progeny.
The purpose has been the search for time modulations of radioactivity by precisely measuring several times per year the half-life of a short lifetime (days) source
instead of measuring over a long period of time the activity of a long lifetime (tens or hundreds of years) source.
As source we have used radon diffused into olive oil: this way we removed
the large fluctuations in the count rate triggered by a temperature gradient at the surface of the volume containing air charged with radon
(a difference of 0.01 K is enough to produce a sizable effect on the count rate).
Our results do not support the
hypothesis of
a possible influence of a component in the solar irradiation affecting the nuclear decay rate.
In particular, we performed four half-life measurements in the period from May 2014 to January 2015 which provide a constant
$^{222}$Rn half-life  with a 8$\cdot$10$^{-5}$  precision. Finally, we improve by one order of magnitude the precision on
$^{222}$Rn half-life providing the value of 3.82146(16)$_{stat}$(4)$_{ syst}$  d.
\section *{Acknowledgments}
We thank Prof. F. Borghesani of Padova University for several enlightening discussions. The Director of LNGS and the staff of the Laboratory are warmly acknowledged for their
support.



%% References


%\bibliographystyle{elsarticle-num}

\begin{thebibliography}{99}
\bibitem{Cur03} M. Curie, Doctoral Dissertation, Sorbonne University, Paris (1903).
\bibitem{Jen09} J.H. Jenkins et al., Astroparticle Physics 32 (2009) 42.
\bibitem{Stu14} P.A. Sturrock et al., The Astrophysical Journal 794:42 (2014).
\bibitem{Bel14} E. Bellotti et al., Astroparticle Physics 61 (2015) 82 (on-line).
\bibitem{Fis11} E. Fischbach et al., arXiv:1106.1470.
\bibitem{Mei14} R.J. de Meijer and S.W. Steyn, arXiv:1409.6969.
\bibitem{Bel12} E. Bellotti et al., Phys. Lett. B 710 (2012) 114.
\bibitem{Bel13} E. Bellotti et al., Phys. Lett. B 720 (2013) 116.
\bibitem{Har12} J.C. Hardy et al., Appl. Radiat. Isot. 70 (2012) 1931.
\bibitem{Ale14} E.N. Alexeyev et al. arXiv:1404.5769.
\bibitem{Fir99} http://www.nucleide.org/DDEP$\_$WG/DDEPdata.htm.
\bibitem{Mon08} M.M B$\acute{e}$ et al., Table of Radionuclides, Vol 4 (2008) 143, Bureau International des Poids et Mesures.
\bibitem{Stu12} P.A. Sturrock et al., Astropart. Phys. 36 (2012) 18.
\bibitem{Ste13} G. Steinitz et al., Geophysical Journal International, 10.1093 (2013).
\bibitem{Ste14} G. Steinitz et al., Journal of Environmental Radioactivity, 134 (2014) 128.
\bibitem{Sol82} UNSCEAR, 1982. Report to the general assembly. Annex. D. Exposure to
radon and thoron and their decay products. P.175, Table 2, United Nations, New York.
\bibitem{Col95} R. Coll$\acute{e}$, Radioactivity $\&$ Radiochemistry 6 (1995) 16.
\bibitem{Chi08} V. Chist$\acute{e}$ and M.M. B$\acute{e}$,   $^{222}$Rn-Comments on evaluation of decay data, in http://www.nucleide.org/DDEP$\_$WG/DDEPdata.htm.
\end{thebibliography}


%\section *{References}

\end{document}